\documentclass[journal]{IEEEtran}
\usepackage{amsmath,amsfonts,bm}
\usepackage[dvips]{graphicx}

\begin{document}

\title{An Efficient Pseudo-Codeword Search Algorithm for Linear Programming Decoding of LDPC Codes}

\author{Michael Chertkov \thanks{M. Chertkov [chertkov@lanl.gov] is in the Theoretical Division and the Center for Nonlinear
Studies, LANL,  Los Alamos, NM 87545, USA.} and Mikhail G. Stepanov
\thanks{M.G. Stepanov [stepanov@math.arizona.edu] is in the
Department of Mathematics, The University of Arizona 617 N. Santa
Rita Ave, P.O. Box 210089 Tucson, AZ 85721-0089 USA, and he is also
a former member and an external affiliate in the Theoretical
Division and the Center for Nonlinear Studies, LANL, Los Alamos, NM
87545, USA}}


\markboth{To appear in IEEE Transactions on Information Theory,
2007}
{Chertkov \& Stepanov: Pseudo-Codeword-Search Algorithm for Linear
Programming Decoding of LDPC Codes}




\maketitle

\begin{abstract}
In Linear Programming (LP) decoding of a Low-Density-Parity-Check
(LDPC) code one minimizes a linear functional, with coefficients
related to log-likelihood ratios, over a relaxation of the polytope
spanned by the codewords \cite{03FWK}. In order to quantify LP
decoding, and thus to describe performance of the error-correction
scheme at moderate and large Signal-to-Noise-Ratios (SNR), it is
important to study the relaxed polytope to understand better its
vertexes, so-called pseudo-codewords, especially those which are
neighbors of the zero codeword. In this manuscript we propose a
technique to heuristically create a list of these neighbors and
their distances. Our pseudo-codeword-search algorithm starts by
randomly choosing the initial configuration of the noise. The
configuration is modified through a discrete number of steps. Each
step consists of two sub-steps. Firstly, one applies an LP decoder
to the noise-configuration deriving a pseudo-codeword. Secondly, one
finds configuration of the noise equidistant from the pseudo
codeword and the zero codeword. The resulting noise configuration is
used as an entry for the next step. The iterations converge rapidly
to a pseudo-codeword neighboring the zero codeword. Repeated many
times, this procedure is characterized by the distribution function
(frequency spectrum) of the pseudo-codeword effective distance. The
effective distance of the coding scheme is approximated by the
shortest distance pseudo-codeword in the spectrum. The efficiency of
the procedure is demonstrated on examples of the Tanner
$[155,64,20]$ code and Margulis $p=7$ and $p=11$ codes ($672$ and
$2640$ bits long respectively) operating over an
Additive-White-Gaussian-Noise (AWGN) channel.
\end{abstract}

\begin{keywords}
LDPC codes, Linear Programming Decoding, Error-floor
\end{keywords}

\section{Introduction I: LDPC codes, Belief Propagation and Linear Programming}
\label{sec:int1}

We consider an LDPC code (cf. Gallager \cite{63Gal}) defined by some
sparse parity-check matrix, $\hat{H}=\{H_{\alpha i};
\alpha=1,\cdots,M; i=1,\cdots,N \}$, of size $M\times N$. A codeword
${\bm\sigma}=\{\sigma_i=0,1; i=1,\dots,N\}$ satisfies all the check
constraints: $\forall \alpha=1,\dots,M$, $\sum_i H_{\alpha
i}\sigma_i=0$ (mod~$2$). We discuss the practical case of finite $N$
and $M$, as opposed to the $N,M\to\infty$ (thermodynamic) limit for
which Shannon capacity theorems were formulated \cite{48Sha}. The
codeword is sent over a noisy channel. To make our consideration
concrete, we consider the AWGN channel. (Notice that all the
discussions and results of this paper can be easily generalized to
other linear channel models.) Corruption of a codeword in the AWGN
channel is described by the following transition probability:
\begin{eqnarray}
{\cal P}({\bm x}|{\bm \sigma})\propto\prod\limits_i
\exp\left[-2s^2(x_i-\sigma_i)^2\right], \label{AWGN}
\end{eqnarray}
where ${\bm x}$ is the signal measured at the channel output and
$2s^2$ is the Signal-to-Noise Ratio (SNR) of the code,  that is
traditionally denoted as $E_c/N_0$. The  Maximum Likelihood (ML)
decoding corresponding to the restoration of the most probable
pre-image ${\bm \sigma}'$ given the output signal ${\bm x}$,
\begin{eqnarray}
 \arg \max\limits_{{\bm \sigma}'} {\cal P}({\bm x}|{\bm \sigma}'),
\label{ML}
\end{eqnarray}
is not feasible in reality since its complexity grows exponentially
with the system size.

LP decoding was introduced by Feldman, Wainwright and Karger
\cite{03FWK} as a computationally efficient approximation to the ML
decoding. Following \cite{03FWK}, let us first notice that
Eq.~(\ref{ML}) can be restated for the AWGN channel as calculating
\begin{eqnarray}
\arg\min_{{\bm\sigma}'\in P}\left(\sum_i(1-2 x_i)\sigma'_i\right),
\label{ML-LP}
\end{eqnarray}
where $P$ is the polytope spanned by the codewords. Looking for
${\bm\sigma}'$ in terms of a linear combination of all possible
codewords of the code, ${\bm \sigma}_v$:
${\bm\sigma}'=\sum_v\lambda_v{\bm \sigma}_v$, where $\lambda_v\geq
0$ and $\sum_v\lambda_v=1$, one finds that ML turns into a linear
optimization problem.  LP decoding proposes to relax the polytope,
expressing ${\bm\sigma}'$ in terms of a linear combination of the
so-called local codewords, i.e. codewords of trivial codes, each
associated with just one check of the original code and all the
variable nodes connected to it. We will come to the formal
definition of the LP decoding \cite{03FWK,03KV,04VK,05VK} later
after discussing the Belief Propagation (BP) decoding of Gallager
\cite{63Gal,68Gal,88Pea,99Mac}.

The belief-propagation (BP), or sum-product, algorithm of Gallager
\cite{63Gal} (see also \cite{68Gal,88Pea,99Mac}) is a popular
iterative scheme often used for decoding of the LDPC codes. For an
idealized code containing no loops (i.e., there is a unique path
connecting any two bits through a sequence of other bits and their
neighboring checks), the sum-product algorithm (with sufficient
number of iterations) is exactly equivalent to the so-called
Maximum-A-Posteriori (MAP) decoding, which is reduced to ML in the
asymptotic limit of infinite SNR. For any realistic code (with
loops), the sum-product algorithm is approximate, and it should
actually be considered as an algorithm for solving iteratively
certain nonlinear equations, called BP equations. The BP equations
minimize the so-called Bethe free energy \cite{05YFW}. (The Bethe
free energy approach originates from a variational methodology
developed in statistical physics \cite{35Bet,51Kik}.) Minimizing the
Bethe free energy, that is a nonlinear function of the
probabilities/beliefs, under the set of linear (compatibility and
normalizability) constraints, is generally a difficult task.

BP decoding becomes LP decoding in the asymptotic limit of infinite
SNR. Indeed in this special limit, the entropy terms in the Bethe
free energy can be neglected and the problem becomes minimization of
a linear function under a set of linear constraints. The similarity
between LP and BP (the latter one being understood as minimizing the
Bethe Free energy \cite{05YFW}) was first noticed in \cite{03FWK}
and it was also discussed in \cite{03KV,04VK,05VK}. Stated in terms
of beliefs, i.e. trial marginal probabilities, LP decoding minimizes
the Bethe self-energy:
\begin{eqnarray}
E=\sum\limits_\alpha\sum\limits_{\sigma_\alpha}
b_\alpha(\sigma_\alpha)\sum\limits_{i\in\alpha}\sigma_i
(1-2x_i)/k_i,\label{LP}
\end{eqnarray}
with respect to beliefs $b_\alpha(\sigma_\alpha)$ and under certain
equality and inequality constraints. Here in Eq.~(\ref{LP}) $k_i$ is
the degree of connectivity of the $i$-th bit; $\sigma_\alpha$ is a
local codeword, $\sigma_\alpha=\{\sigma_i|i\in\alpha,\sum_i
H_{\alpha i}\sigma_i=0\mbox{ (mod~$2$)}\}$, associated with the
check $\alpha$. The equality constraints are of two types,
normalization constraints (beliefs, as probabilities, should sum to
one) and compatibility constraints
\begin{eqnarray}
&& \forall\;\alpha: \ \ \sum\limits_{\sigma_\alpha}b_\alpha(\sigma_\alpha)=1,\label{norm}\\
&& \forall\; i\ \forall\;\alpha\ni i:\ \
b_i(\sigma_i)=\sum\limits_{\sigma_\alpha\setminus\sigma_i}b_\alpha(\sigma_\alpha),\label{comp}
\end{eqnarray}
respectively  where $b_i(\sigma_i)$ is the belief (trial marginal
probability) to find bit $i$ in the state $\sigma_i$, and the check
belief, $b_\alpha({\bm\sigma}_\alpha)$, stands for the trial
marginal probability of finding bits, which are neighbors of the
check $\alpha$, in the state ${\bm\sigma}_\alpha$. Also, all the
beliefs, as probabilities, should be non-negative and smaller than
or equal to unity. Thus there is the additional set of the
inequality constraints:
\begin{eqnarray}
0\leq b_i(\sigma_i),b_\alpha(\sigma_\alpha)\leq 1. \label{ineq}
\end{eqnarray}

\section{Introduction II: Pseudo codewords, Frame Error Rate and
effective distance} \label{sec:int2}

As it was shown in \cite{03FWK} the LP decoding has ML certificate,
i.e. if the pseudo-codeword obtained by the LP decoder has only
integral entries then it must be a codeword,  in fact it is the
codeword given back by ML decoder. If LP decoding does not decode to
a correct codeword then it usually yields a non-codeword
pseudo-codeword with some number of non-integers among the beliefs
$b_i$ and $b_\alpha$. These configurations can be interpreted as
mixed state configurations consisting of a probabilistic mixture of
local codewords.

An important characteristic of the decoding performance is Frame
Error Rate (FER) calculating the probability of decoding failure.
FER decreases as SNR increases. The form of this dependence gives an
ultimate description of the coding performance. Any decoding to a
non-codeword pseudo-codeword is a failure. Decoding to a codeword
can also be a failure, which counts as a failure under ML decoding.
For large SNR, i.e. in the so-called error-floor domain, splitting
of the two (FER vs SNR) curves, representing the ML decoding and an
approximate decoding (say LP decoding) is due to pseudo-codewords
\cite{04Ric}. The actual asymptotics of the two curves for the AWGN
channel are $\mbox{FER}_{\mbox{\scriptsize ML}}\sim \exp(-
d_{\mbox{\scriptsize ML}}\cdot s^2/2)$ and
$\mbox{FER}_{\mbox{\scriptsize LP}}\sim \exp(-d_{\mbox{\scriptsize
LP}}\cdot s^2/2)$, where $d_{\mbox{\scriptsize ML}}$ is the
so-called Hamming distance of the code and the $d_{\mbox{\scriptsize
LP}}$ is the effective distance of the code, specific for the LP
decoding. The LP error-floor asymptotic is normally shallower than
the ML one, $d_{\mbox{\scriptsize LP}}<d_{\mbox{\scriptsize ML}}$.
The error floor can start at relatively low values of FER,
unaccessible for Monte-Carlo simulations. This emphasizes importance
of the pseudo-codewords analysis.

For a generic linear code performed over a symmetric channel, it is
easy to show that the FER is invariant under the change of the
original codeword (sent into the channel). Therefore, for the
purpose of FER evaluation, it is sufficient to analyze the statistic
exclusively for the case of one codeword, and the choice of zero
codeword is natural. Then calculating the effective distance of a
code, one makes an assumption that there exists a special
configuration (or maybe a few special configurations) of the noise,
instantons according to the terminology of \cite{05SCCV}, describing
the large SNR error-floor asymptotic for FER. Suppose a pseudo
codeword, $\tilde{\bm \sigma}=\{\tilde{\sigma_i}=b_i(1);\
i=1,\dots,N\}$, corresponding to the most damaging configuration of
the noise (instanton), ${\bm x}_{\mbox{\scriptsize inst}}$, is
found. Then finding the instanton configuration itself (i.e.
respective configuration of the noise) is equivalent to maximizing
the transition probability (\ref{AWGN}) with respect to the noise
field, ${\bm x}$, taken at ${\bm \sigma}=0$ under the condition that
the self-energy calculated for the pseudo-codeword in the given
noise field ${\bm x}$ is zero (i.e. it is equal to the value of the
self energy for the zero code word). The resulting expression for
the optimal configuration of the noise (instanton) is
\begin{eqnarray}
 {\bm x}_{\mbox{\scriptsize inst}}=\frac{\tilde{\bm \sigma}}{2}
 \frac{\sum_i\tilde{\sigma}_i}{\sum_i\tilde{\sigma}_i^2},
 \label{inst}
\end{eqnarray}
and the respective effective distance is
\begin{eqnarray}
d_{\mbox{\scriptsize
LP}}=\frac{\left(\sum_i\tilde{\sigma}_i\right)^2}
{\sum_i\tilde{\sigma}_i^2}.\label{dLP}
\end{eqnarray}
This definition of the effective distance was first described in
\cite{01FKKR},  with the first applications of this formula to the
LP decoding discussed in \cite{03KV} and \cite{05VK}. Note also that
Eqs.~(\ref{inst},\ref{dLP}) are reminiscent of the formulas derived
by Wiberg and co-authors in \cite{95WLK} and \cite{96Wib}, in the
context of the computational tree analysis applied to iterative
decoding with a finite number of iterations.

\begin{figure}[t]
\centerline{\includegraphics[width=3.1in]{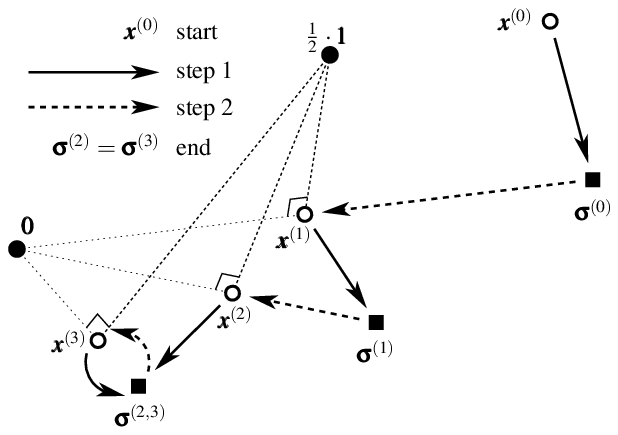}}
\caption{Schematic illustration of the pseudo-codeword-search
algorithm. This example terminates at $k_*=3$. The point ${\bm
1}/2=(1,\cdots,1)/2$ is shown to illustrate that if one draws a
straight line through ${\bm 1}/2$, such that it is perpendicular to
the straight line connecting ${\bm 0}=(0,\cdots,0)$ and ${\bm
\sigma}^{(k)}$, then the straight line must go
$\varepsilon$-approximately through ${\bm x}^{(k+1)}$. [We are
thankful to Referee A for making this useful ${\bm 1}/2$-related
observation.]} \label{fig:scheme}
\end{figure}

\section{Searching for pseudo-codewords}

In this Section, we turn directly to describing an algorithm which
allows one to find efficiently pseudo-codewords of an LDPC code
performing over AWGN channel and decoded by LP. Once the algorithm
is formulated, its relation to the introductory material, as well as
partial justification and motivation will become clear.

\underline{\bf The Pseudo-codeword search algorithm:}
\begin{itemize}
\item{\bf Start:} Initiate a starting configuration of the noise, $
{\bm x}^{(0)}$. Noise measures a deviation from the zero codeword
and it should be sufficiently large to guarantee convergence of LP
to a pseudo-codeword different from the zero codeword.

\item{\bf Step 1:} The LP decoder finds the closest pseudo-codeword,
${\bm \sigma}^{(k)}$, for the given configuration of the noise
\begin{eqnarray} &&
 \!\!\!\!\{b_i^{(\mbox{\scriptsize
LP},k)}(\sigma_i),b_\alpha^{(\mbox{\scriptsize
LP},k)}(\sigma_\alpha)\}\nonumber\\
 && =\arg\!\!\!\!\!\!\!\!
 \!\!\!\!\min\limits_
 {\{b_i(\sigma_i), b_\alpha(\sigma_\alpha)\}}\!\!\Biggl\{E\left({\bm x}^{(k)};
 \{b_i(\sigma_i), b_\alpha(\sigma_\alpha)\}\right)\nonumber\\
 &&  \mbox{satisfying Eqs.~(\ref{norm},\ref{comp},\ref{ineq})}\Biggr\},
 \nonumber\\ &&
 \sigma_i^{(k)}=b_i^{(\mbox{\scriptsize
LP},k)}(1),\nonumber
\end{eqnarray}
where the self-energy is defined according to Eq.~(\ref{LP}). In the
case of degeneracy one picks any of the closest pseudo-codewords.

\item{\bf Step 2:} Find ${\bm y}^{(k)}$, the
weighted median in the noise space between the pseudo codeword,
${\bm \sigma}^{(k)}$, and the zero codeword:
\begin{eqnarray}
 {\bm y}^{(k)}=\frac{{\bm \sigma}^{(k)}}{2}\frac{\sum_i\sigma_i^{(k)}}
 {\sum_i\big(\sigma_i^{(k)}\big)^2}.
\nonumber
\end{eqnarray}

\item{\bf Step 3:} If ${\bm y}^{(k)}={\bm y}^{(k-1)}$, then $k_*=k$ and the algorithm
terminates. Otherwise go to {\bf Step 2} assigning ${\bm
x}^{(k+1)}={\bm y}^{(k)}+\varepsilon$ for some very small
$\varepsilon$. ($+\varepsilon$ prevents decoding into the zero
codeword, keeping the result of decoding within the erroneous
domain.)
\end{itemize}

{\bf ${\bm y}^{(k_*)}$ is the output}  configuration of the noise
that belongs to the error-surface surrounding the zero codeword.
(The error-surface separates the domain of correct LP decisions from
the domain of incorrect LP decisions.) Moreover, locally,  i.e. for
the given part of the error-surface equidistant from the zero
codeword and the pseudo codeword ${\bm \sigma}^{(k_*)}$, ${\bm
y}^{(k_*)}$ is the nearest point of the error-surface to the zero
codeword.

The algorithm is schematically illustrated in Fig.~\ref{fig:scheme}.
We repeat the algorithm many times picking the initial noise
configuration randomly,  however guaranteeing that it would be
sufficiently far from the zero codeword so that the result of the LP
decoding (first step of the algorithm) is a pseudo-codeword distinct
from the zero codeword. Our simulations (see discussions below) show
that the algorithm converges,  and it does so in a relatively small
number of iterations. The convergence of the algorithm is translated
into the statement that the effective distance between ${\bm
x}^{(n)}$ and the zero codeword does not increase,  but typically
decreases, with iterations. Once the algorithm converges the
resulting pseudo-codeword belongs to the error-surface. This
observation was tested by shifting the instanton configuration of
the noise correspondent to the pseudo-codeword towards the zero
codeword and observing that the result of decoding is the zero
codeword. The effective distance of the coding scheme is
approximated by
\begin{eqnarray}
d_{\mbox{\scriptsize LP}}\approx
\min\Biggl\{\frac{\left(\sum_i\sigma_i^{(k_*)}\right)^2}
{\sum_i\big(\sigma_i^{(k_*)}\big)^2}\Biggr\},
 \label{d-frac}
\end{eqnarray}
where the minimum is taken over multiple evaluations of the
algorithm. It is not guaranteed that the noise configuration with
the lowest possible (of all the pseudo-codewords within the decoding
scheme) distance is found after multiple evaluations of the
algorithm. Also, we do not have a formal proof of the fact that,
beginning with a random ${\bm x}^{(0)}$, our algorithm explores the
entire phase space of all pseudo-codewords on the error-surface.
However our working conjecture is that the rhs of Eq.~(\ref{d-frac})
gives a very tight (if the number of attempts is sufficient) upper
bound on the actual effective distance of the coding scheme.

\section{Examples}

In this Section, we demonstrate the power of the simple procedure
explained in the previous Section by considering three popular
examples of relatively long regular LDPC codes.

\subsection{The Tanner $[155,64,20]$ code of \cite{01TSF}}

\begin{figure}[t]
\centerline{\includegraphics[width=3.1in]{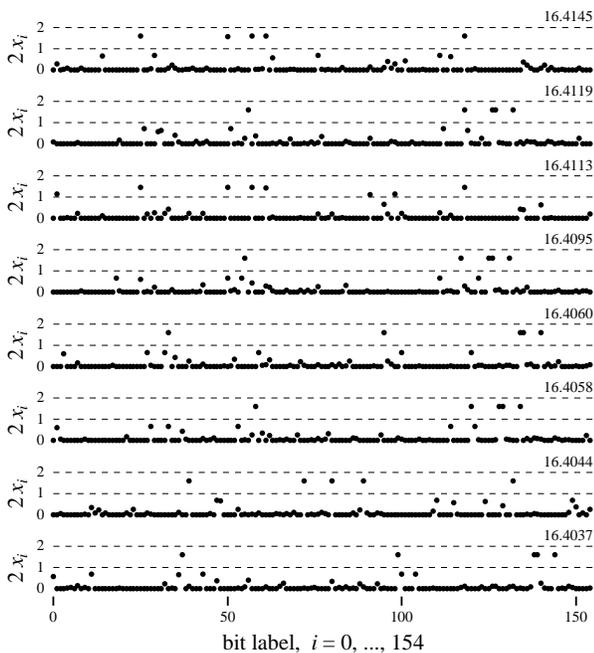}}
\caption{The $8$ lowest configurations found by the
pseudo-codeword-search algorithm for the $[155,64,20]$ code. The
typical number of evaluations required to reach a stopping point is
$5\div 15$. } \label{fig:Tan1}
\end{figure}

\begin{figure}[t]
\centerline{\includegraphics[width=3.1in]{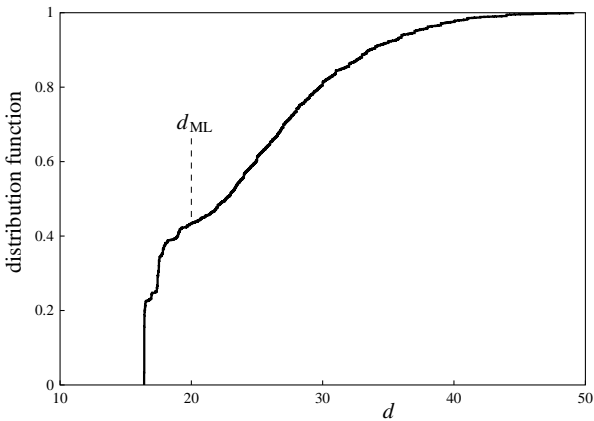}}
\caption{The frequency spectrum (distribution function) of the
effective distance constructed from 3,000 attempts of our
pseudo-codeword search algorithm for the $[155,64,20]$ code. }
\label{fig:Tan2}
\end{figure}

For this code $N=155$ and $M=93$. The Hamming distance of the code
is known to be $d_{\mbox{\scriptsize ML}}=20$. The authors of
\cite{03KV} reported a pseudo codeword with $d=16.406$.   The lowest
effective distance configuration found as a result of our search
procedure is $d_{\mbox{\scriptsize LP}} \approx 16.4037$. These two,
and some number of other lower lying (in the sense of their
effective distance) configurations, are shown in
Fig.~\ref{fig:Tan1}. The resulting frequency spectra (derived from
$3,000$ evaluations of the pseudo-codeword-search algorithm) is
shown in Fig.~\ref{fig:Tan2}. Notice that the pseudo-weight spectrum
gap, defined as the difference between the pseudo-weight of the
non-codeword minimal pseudo-codeword with smallest pseudo-weight and
the minimum distance \cite{05VSKTV}, is negative for the code,
$\approx -3.5963$. Thus the LP decoding performance is strictly
worse than the ML decoding performance for SNR $\to\infty$.

\subsection{The Margulis code \cite{82Mar} with $p=7$}

\begin{figure}[t]
\centerline{\includegraphics[width=3.1in]{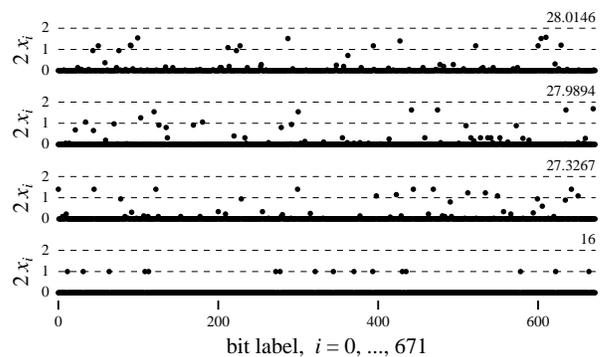}}
\caption{The $4$ lowest noise configuration found by our
pseudo-codeword search algorithm for the Margulis $p=7$ code of
\cite{82Mar}. The typical number of evaluations required to reach a
stopping point is in between $10$ and $20$. } \label{fig:M7_1}
\end{figure}

\begin{figure}[t]
\centerline{\includegraphics[width=3.1in]{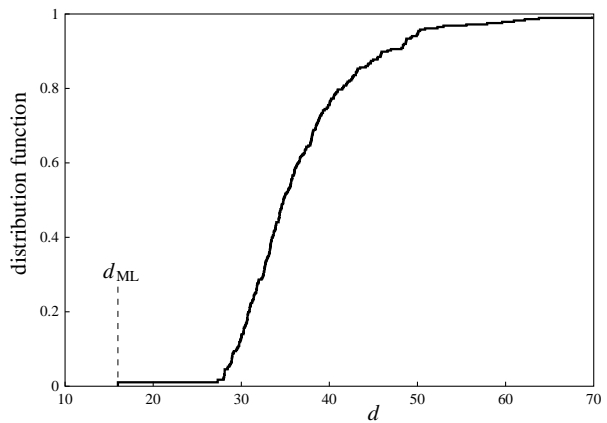}}
\caption{The frequency spectrum (distribution function)  of the
effective distance found through multiple attempts of the
pseudo-codeword-search algorithm for the Margulis $p=7$ code. The
figure is built on $~250$ evaluations of the pseudo-codeword-search
algorithm. } \label{fig:M7_2}
\end{figure}

This code has $N=2\cdot M=672$ bits. The set of four noise
configurations with the lowest effective distance found by the
pseudo-codeword-search algorithm for the code is shown in
Fig.~\ref{fig:M7_1}. The lowest configuration decodes into a
codeword with the Hamming distance $16$.  A large gap separates this
configuration from the next lowest configuration corresponding to a
pseudo-codeword that is not a codeword. Since the pseudo-weight
spectrum gap is positive in this case, the LP decoding approaches
the ML decoding performance for SNR $\to\infty$. The frequency
spectra, characterizing the performance of the
pseudo-codeword-search algorithm for this code, is shown in
Fig.~\ref{fig:M7_2}.

\subsection{The Margulis code \cite{82Mar} with $p=11$}

\begin{figure}[h]
\centerline{\includegraphics[width=3.1in]{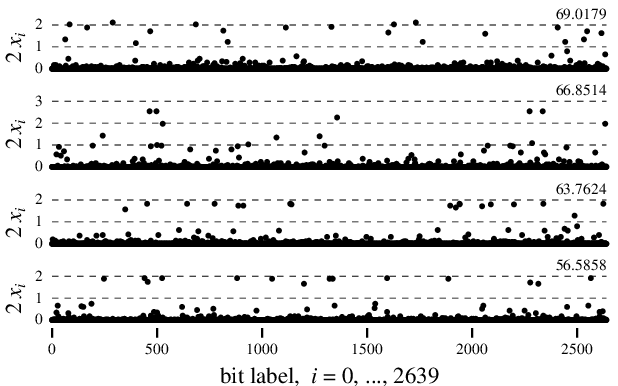}}
\caption{The $4$ lowest noise configurations found by our
pseudo-codeword search algorithm for the Margulis $p=11$ code of
\cite{82Mar}. The typical number of the pseudo-codeword-search
iterations required to reach a stopping point is in between $30$ and
$60$. } \label{fig:M11}
\end{figure}

This code is $N=2\cdot M=2640$ bits long. We have a relatively small
number of configurations ($30$) here because it takes much longer to
execute the LP decoding in this case. Some $30$ to $60$ steps of the
pseudo-codeword search are required for a typical realization of the
algorithm to reach a stopping point. The four lowest configurations
are shown in Fig.~\ref{fig:M11}. Obviously, with limited statistics
one cannot claim that the noise configuration with the lowest
possible effective length has been found. All stopping point
configurations found here correspond to pseudo codewords. (The
Hamming distance for this code is not known, while the pessimistic
upper bound mentioned in \cite{02MP} is $220$.)

\section{Conclusions and Discussions}

Let us discuss the utility of the pseudo-codeword search algorithm
proposed in this manuscript. The algorithm gives an efficient way of
describing the LP decoding polytope and the pseudo-codeword spectra
of the code.  It approximates the pseudo-codeword and the respective
noise configuration on the error-surface surrounding the zero
codeword, corresponding to the shortest effective distance of the
code. Our test shows that the algorithm converges very rapidly.
(Even for the $2640$ bits code,  the longest code we considered, it
typically takes only $30$ to $60$ steps of the pseudo-codeword
search algorithm to converge.) As already mentioned,  this procedure
applies to any linear channel. One only needs to make modifications
in Eqs.~(\ref{inst},\ref{dLP},\ref{d-frac}) and also in the basic
equation of Step 2.

One would obviously be interested in extending the pseudo-codeword
search algorithm to other decodings, e.g. to find the effective
minimal distance of the sum-product decoding. We observed, however,
that a naive extension of this procedure does not work. The very
special feature of the LP-case is that the noise configuration found
as a weighted median of the zero codeword and a pseudo codeword
($+\varepsilon$, as in the Step 3 of the pseudo-codeword search
algorithm)  {\it is not} decoded into the zero codeword. This allows
us to proceed with the search algorithm always decreasing the
effective distance or at least keeping it constant. It is not yet
clear if this key feature of the LP decoding is extendable
(hopefully with some modification of the weighted median procedure)
to iterative decoding. This question requires further investigation.

Even though the direct attempt to extend the LP-based
pseudo-codewords-search algorithm to the sum-product decoding
failed, we still found an indirect way of using these LP results to
analyze the sum-product decoding. The most damaging configuration of
the noise found within the pseudo-codeword-search procedure becomes
a very good entry point for the instanton-amoeba method of
\cite{05SCCV}, designed for finding instanton configurations (most
damaging configurations of the noise) for the case of the standard
iterative decoding. This hybrid method works well, sometimes
resulting in the discovery of pseudo-codewords (of the respective
iterative scheme) with impressively small effective distance. We
attribute this fact to the close relation existing between the LP
decoding and the BP decoding \cite{03FWK,03KV,04VK,05VK}. Some
preliminary results of this hybrid analysis are discussed in
\cite{06SC}. Summarizing, the LP-based pseudo-codeword search
algorithm, complemented and extended by the instanton-amoeba method
of \cite{05SCCV}, provides an efficient practical tool for the
analysis of effective distances, most damaging configurations of the
noise (instantons) describing the error-floor, and their frequency
spectra for an arbitrary LDPC code performing over a linear channel
and decoded by LP decoding or iteratively.

After the original version of the manuscript was submitted for
publication,  we have learned about some important new results
concerning reducing complexity of LP-decoding \cite{06TS,06VK}. It
is also appropriate to mention here the most recent publications
exploring possibilities of LP-decoding improvement
\cite{06DW,06CCc}. These new techniques and ideas combined with the
pseudo-codeword-search algorithm  open interesting new opportunities
for exploring and improving decoding schemes of even longer LDPC
codes.

\section{Acknowledgements}

The authors acknowledge very useful, inspiring and fruitful
discussions with Vladimir Chernyak, Ralf Koetter, Olgica Milenkovich
and Bane Vasic. This work was carried out under the auspices of the
National Nuclear Security Administration of the U.S. Department of
Energy at Los Alamos National Laboratory under Contract No.
DE-AC52-06NA25396.

\end{document}